\documentclass[12pt]{article}
\pdfoutput =1
\usepackage{float} 
\usepackage{amsmath}
\usepackage{slashed}
\usepackage{amsfonts}   
\usepackage{amssymb}

\begin{document}
\date{}
\title{{\bf{\Large  Analytic integrability for holographic duals with $ J\bar{T} $ deformations}}}
\author{
 {\bf {\normalsize Dibakar Roychowdhury}$
$\thanks{E-mail:  dibakarphys@gmail.com, dibakarfph@iitr.ac.in}}\\
 {\normalsize  Department of Physics, Indian Institute of Technology Roorkee,}\\
  {\normalsize Roorkee 247667, Uttarakhand, India}
\\[0.3cm]
}

\maketitle
\begin{abstract}
We probe warped BTZ $ \times S^3 $ geometry with various string solitons and explore the classical integrability criteria of the associated phase space configurations using Kovacic's algorithm. We consider consistent truncation of the parent sigma model into one dimension and obtain the corresponding normal variational equations (NVE).  Two specific examples have been considered where the sigma model is reduced over the subspace of the full target space geometry. In both examples, NVEs are found to possess Liouvillian form of solutions which ensures the classical integrability of the associated phase space dynamics. We address similar issues for the finite temperature counterpart of the duality, where we analyse the classical phase space of the string soliton probing warped BTZ black string geometry. Our analysis reveals a clear compatibility between normal variational equations and the rules set by the Kovacic's criteria. This ensures the classical integrability of the parent sigma model for the finite temperature extension of the duality conjecture.
\end{abstract}
\section{Overview and Motivation}
Our recent understanding on gauge/string correspondence has revealed a remarkable duality between string sigma models on warped $ AdS_3\times S^3 $ \cite{Anninos:2008fx}-\cite{Apolo:2019zai} and solvable \emph{irrelevant} deformations \cite{Zamolodchikov:2004ce}-\cite{Frolov:2019xzi} in 2D CFTs \cite{Apolo:2018qpq}-\cite{Apolo:2019yfj}. The stringy side of the duality is obtained via \emph{marginal} deformations of the $ AdS_3 \times S^3 $ world-sheet theory by suitable choice of vertex operators\footnote{From the point of view of the boundary CFT$ _2 $, the vertex operator (\ref{e1}) acts as a source of \emph{single} trace $ J\bar{T} $ deformation $ \sim \mu \int d^2x  ~\mathcal{A}^{3}(x , \bar{x})$ to the boundary Lagrangian \cite{Chakraborty:2018vja}.} those are observables with scaling dimension (1, 2) in the boundary CFT$ _2 $ \cite{Apolo:2019yfj},
\begin{eqnarray}
\mathcal{A}^{a}(x , \bar{x})=\frac{1}{2}\int d^2z (\partial_x \bar{J}(\bar{x}, \bar{z})\partial_{\bar{x}}+2 \partial^2_{\bar{x}}\bar{J}(\bar{x},\bar{z}))\Phi_{1}(x,\bar{x};z, \bar{z})k^a (z)
\label{e1}
\end{eqnarray}
where $x(=\varphi +t)$ and $ \bar{x} (=\varphi -t)$ are respectively the left and right moving coordinates associated with the dual CFT. Here, $\bar{J}(\bar{x}, \bar{z})$ is a linear combination of $ SL(2,R)_R $ world-sheet currents ($ \bar{j}^- (\bar{z}) $), $ k^a $ stands for an $ SU(2)_L $ world-sheet current and $ \Phi_{1}(x,\bar{x};z, \bar{z}) $ correponds to bulk to boundary propagator \cite{Kutasov:1999xu}.

The corresponding \emph{marginal} current current deformation (on the world-sheet/ spacetime  CFT$ _2 $) dual to $ J \bar{T} $ deformations of the boundary CFT$ _2 $ \cite{Chakraborty:2018vja},
\begin{eqnarray}
\Delta S_{\lambda}=\frac{2 \lambda}{\mathsf{k}}\int d^2z~ k^3 (z)\bar{j}^- (\bar{z})
\end{eqnarray}
results in the target space fields that could be formally expressed as\footnote{This also goes under the name of zero temperature/ warped massless solution \cite{Apolo:2019yfj}.} \cite{Apolo:2019yfj},
\begin{eqnarray}
\label{e2}
ds^2&=&\frac{dr^2}{4r^2}+r(d\varphi^{2}-dt^2)+\lambda r (d \varphi +d t)(d \chi + \cos\theta d\psi)+d\Omega_{3}^2\\
d\Omega_{3}^2&=&\frac{1}{4}\left((d \chi + \cos\theta d\psi)^2 +d\theta^2 +\sin^2\theta d\psi^2\right) \\
\mathfrak{B}_{2}&=&\frac{1}{4}\left(\cos\theta d\psi\wedge d\chi +4r d\varphi \wedge dt \right) -\frac{\lambda r}{2}(d \varphi +dt)\wedge (d \chi + \cos\theta d\psi)\\
e^{-2\Phi}&=&1
\label{e4}
\end{eqnarray}
where we set the AdS length scale, $ l=1 $. Here, $ \lambda $ is the deformation parameter and $ \mathsf{k} $ is the number of $ NS5 $ branes wrapping the internal space which can be identified with the central charge of the seed theory in the dual symmetric product CFT$_2$ \cite{Apolo:2019yfj}. The resulting background solutions (\ref{e2})-(\ref{e4}) could also be obtained from their undeformed cousins \cite{Apolo:2019yfj} via TsT transformations. In other words, the marginal deformation on world-sheet fields is equivalent of applying a TsT on $ AdS_3 \times S^3 $ solutions in type IIB supergravity \cite{Araujo:2018rho}. 

The dual field theory counterpart of (\ref{e2})-(\ref{e4}) has been identified with 2D conformal field theory (at zero temperature) with \emph{irrelevant} deformation of the type,
\begin{eqnarray}
\mathcal{O}_{J\bar{T}}=J_{x}\bar{T}-\bar{J}T_{x \bar{x}}
\label{e5}
\end{eqnarray}
where, $ \bar{T}=T_{\bar{x}\bar{x}} $ is the right moving component of the stress tensor together with $ \bar{J}=J_{\bar{x}} $ as the right moving component of the  $ U(1)$ current \cite{Guica:2017lia}. The deformation of the above type (\ref{e5}) is found preserve the $ SL(2,R)_{L}\times U(1)_R $ subgroup of the original CFT$ _{2} $ \cite{Bzowski:2018pcy} and triggers a RG flow from IR CFT$ _2 $ to \emph{non local} QFTs at UV those are relevant from the perspective of the extremal Kerr/CFT correspondence \cite{Guica:2008mu}.

Recently, our understanding on the above correspondence has been lifted to its finite temperature realization by matching the stringy spectrum on the finite temperature background with its 2D CFT counterpart \cite{Apolo:2019yfj}. On the stringy side of the correspondence, the finite temperature background has been obtained by applying TsT transformations on BTZ black strings thereby lifting it to warped BTZ black strings \cite{Azeyanagi:2012zd}. At the level of the world-sheet theory, such deformations correspond to an action of instantaneous current-current deformation sourced due to antisymmetric product of two Noether currents.

Given the gauge/string duality as depicted above, the purpose of the present paper is to pose the following question namely whether the world-sheet theory dual to $ J\bar{T} $ deformed field theories is integrable or not - at least for the bosonic sector. The traditional way of addressing such questions comes through the systematic construction of Lax pairs \cite{Orlando:2019rjg} which is in general a non trivial task to perform. Therefore, in this paper we choose work with an alternate path which is to adopt certain analytic tools that would eventually help us to disprove integrability rather than proving it. While on one hand this looks like a tractable project to start with, on the other hand, there appears to be a bunch of non-linear partial differential equations popping up during the consistent reduction of the sigma model over the deformed target space geometry. These equations could be studied both analytically as well as numerically. Based on certain standard prescriptions (that we elaborate below), our strategy would be to check whether any of these truncations admits non integrability and if found positive then we would conclude that the parent sigma model is non-integrable. On the other hand, in case of negative results we would conclude that the parent sigma model that we had started with is classically integrable.

We probe warped $ AdS_3 \times S^3 $ geometries with various string solitons and explore the corresponding phase space dynamics by means of analytic tools whose `logical' foundation is based on the so called Kovacic's algorithm \cite{kovacic1}-\cite{kovacic2}. The algorithm offers an useful way of classifying the Hamiltonian phase space dynamics based on its solvability and has been applied with a remarkable success in order to extract informations regarding integrability or non integrability of the string solitons in the context of gauge/string duality \cite{Basu:2011fw}-\cite{Nunez:2018qcj}. 

The recipe to check analytic (non)integrability comprises of the following steps - typically one chooses to work with an invariant plane \cite{Stepanchuk:2012xi} in the phase space and consider fluctuations ($ \eta (\tau) $) normal to this plane. At leading order, these fluctuations satisfy second order linear differential equations\footnote{Also known as normal variational equations (NVE) \cite{Stepanchuk:2012xi}.} of the form \cite{Stepanchuk:2012xi},
\begin{eqnarray}
\mathcal{A}(\tau)\ddot{\eta}+\mathcal{B}(\tau)\dot{\eta}+\mathcal{C}(\tau)\eta = 0
\label{e8}
\end{eqnarray}
where the coefficients $ \mathcal{A} $ , $ \mathcal{B} $ and $ \mathcal{C} $ are rational polynomials. For \emph{integrable} classical Hamiltonian systems, solutions corresponding to NVE (\ref{e8}) could be expressed as simple algebraic polynomials/exponentials/logarithmic functions known as \emph{Liouvillian} solutions \cite{Basu:2011fw}-\cite{Stepanchuk:2012xi}. Kovacic's algorithm comprises of systematic steps to check whether such solutions are in fact pertinent to (\ref{e8}).  If any particular (phase space) configuration does not fit with the framework set by the algorithm, then the associated Hamiltonian dynamics is classified as \emph{non-integrable}. Therefore, the algorithm essentially allows us for a case by case study of various phase space configurations to pin down those which are non-integrable.

In order to understand the working principle behind the algorithm, it is useful to consider the following transformation \cite{Nunez:2018qcj},
\begin{eqnarray}
\eta (\tau)=e^{\int w(\tau)-\frac{\tilde{\mathcal{B}}(\tau)}{2}}~;~\tilde{\mathcal{B}}=\frac{\mathcal{B}}{\mathcal{A}}
\end{eqnarray}
which rewrites (\ref{e8}) as,
\begin{eqnarray}
\dot{w}(\tau)+w^2(\tau)=V(\tau)=\frac{2\tilde{\mathcal{B}}'+\tilde{\mathcal{B}}^2-4\tilde{\mathcal{C}}}{4}~;~\tilde{\mathcal{C}}=\frac{\mathcal{C}}{\mathcal{A}}.
\label{e10}
\end{eqnarray}
As per the rules set by the algorithm, the Liouvillian form of solutions is guaranteed iff $ w(\tau) $ is a rational polynomial of degree 1, 2, 4, 6 or 12. This result comes from the general group of invariance associated with the corresponding space of solutions which is a subgroup of $ SL(2,C) $ \cite{kovacic1}-\cite{kovacic2}, \cite{Roychowdhury:2017vdo}, \cite{Nunez:2018qcj}. Therefore, as per the algorithm, the analytic integrability is ensured once $ w(\tau) $ belongs to one of the above polynomial classes.

In order to claim integrability or non-integrability of the string soliton probing warped $ BTZ\times S^3 $, we therefore strictly follow the steps as mentioned above. We perform our computations both for the zero as well as finite temperature examples of the duality. We first consistently reduce the sigma model over different subspaces of the full target space geometry and obtain the corresponding normal variational equations those are compatible with the various other constraints of the system. Finally, we examine whether the resulting phase space configurations are in fact compatible with the rules set by the algorithm. This is achieved through computation of the corresponding potential function $ V(\tau) $ and thereby obtaining the solution ($ w(\tau) $) corresponding to (\ref{e10}). Finally, we conclude in Section 3.
\section{Analysis and results}
We start our analysis considering the parent sigma model in conformal gauge,
\begin{eqnarray}
\mathfrak{S}_{P}=\frac{1}{4 \pi \alpha'}\int d\tau d\sigma (\eta^{\alpha \beta}\mathfrak{G}_{MN}+\varepsilon^{\alpha \beta}\mathfrak{B}_{MN})\partial_{\alpha}X^M \partial_{\beta}X^N=\frac{1}{4 \pi \alpha'}\int d\tau d\sigma ~ \mathfrak{L}_P
\label{e11}
\end{eqnarray}
where ($ \tau , \sigma $) are the so called world-sheet coordinates, $ X^M $s are the target space coordinates and $ (\alpha')^{-1} $ measures the tension of the string soliton. Moreover, we define $\varepsilon^{\alpha \beta}  $ as the 2D Levi-Civita symbol such that $ \varepsilon^{\tau \sigma}=-\varepsilon^{\sigma \tau}=1 $.
\subsection{Sigma models on warped massless BTZ$ \times S^3 $ }
Using (\ref{e2})-(\ref{e4}), the sigma model Lagrangian (\ref{e11}) could be formally expressed as,
\begin{eqnarray}
\mathfrak{L}_P=r(\dot{t}^2 - t'^2)+\frac{1}{4r^2}(r'^2 - \dot{r}^2) +r (\varphi'^2 - \dot{\varphi}^2)+\frac{1}{4}(\chi'^2 - \dot{\chi}^2)+\frac{1}{4}(\theta'^2 - \dot{\theta}^2)\nonumber\\
+\frac{1}{4}(\psi'^2 - \dot{\psi}^2)+2 \lambda r (t' \chi' -\dot{t}\dot{\chi})+ 2 \lambda r \cos\theta (t' \psi' - \dot{t}\dot{\psi})+2 \lambda r (\varphi' \chi' -\dot{\varphi}\dot{\chi})\nonumber\\
+ 2 \lambda r \cos\theta (\varphi' \psi' -\dot{\varphi}\dot{\psi})+ \cos\theta (\psi' \chi' - \dot{\psi}\dot{\chi})+\frac{\cos\theta}{2}(\dot{\psi} \chi' - \dot{\chi} \psi')+2r(\dot{\varphi}t' -\dot{t}\varphi')\nonumber\\
- \lambda r (\dot{\varphi}\chi' - \dot{\chi}\varphi')-\lambda r \cos\theta (\dot{\varphi}\psi' -\dot{\psi}\varphi')-\lambda r (\dot{t}\chi' - \dot{\chi}t')-\lambda r \cos \theta (\dot{t}\psi' -\dot{\psi}t')
\label{e12}
\end{eqnarray}
where dot corresponds to derivative w.r.t. $ \tau $ and prime stands for the derivative w.r.t. $ \sigma $.

Given (\ref{e12}), the conserved charge densities associated with the 2D sigma model follow immediately,
\begin{eqnarray}
\mathcal{E}&=&\frac{\delta \mathfrak{L}_P}{\delta \dot{t}}=2 r \dot{t}-2 \lambda r \dot{\chi}-2 \lambda r \cos\theta \dot{\psi}+2 r \varphi' - \lambda r \chi' - \lambda r \cos\theta \psi'\\
\mathcal{P}_{\varphi}&=&\frac{\delta \mathfrak{L}_P}{\delta \dot{\varphi}}=-2 r \dot{\varphi}-2 \lambda r \dot{\chi}- 2 \lambda r \cos\theta \dot{\psi}- 2 r t' - \lambda r \chi' - \lambda r \cos\theta \psi'\\
\mathcal{J}_{\chi}&=&\frac{\delta \mathfrak{L}_P}{\delta \dot{\chi}}=-\frac{\dot{\chi}}{2}- 2 \lambda r \dot{t}- 2 \lambda r \dot{\varphi}-\cos\theta \dot{\psi}-\frac{\cos\theta}{2}\psi' + \lambda r \varphi' + \lambda r t'\\
\mathcal{J}_{\psi}&=&\frac{\delta \mathfrak{L}_P}{\delta \dot{\psi}}=-\frac{\dot{\psi}}{2}-2 \lambda r \cos\theta \dot{t}-2 \lambda r \cos\theta \dot{\varphi}-\cos\theta \dot{\chi}+\frac{\cos\theta}{2}\chi' + \lambda r \cos\theta \varphi' +\lambda r \cos\theta t'.\nonumber\\
\end{eqnarray}

Finally, we note down the Virasoro constraints,
\begin{eqnarray}
T_{\tau \tau}=-r(t'^2 + \dot{t}^2)+\frac{1}{4r^2}(r'^2 + \dot{r}^2)+r (\varphi'^2 + \dot{\varphi}^2)+\frac{1}{4}(\chi'^2 + \dot{\chi}^2)+\frac{1}{4}(\theta'^2 + \dot{\theta}^2)\nonumber\\
+\frac{1}{4}(\psi'^2 + \dot{\psi}^2)+2 \lambda r (t' \chi' +\dot{t}\dot{\chi})+ 2 \lambda r \cos\theta (t' \psi' + \dot{t}\dot{\psi})+2 \lambda r (\varphi' \chi' +\dot{\varphi}\dot{\chi})\nonumber\\
+ 2 \lambda r \cos\theta (\varphi' \psi' +\dot{\varphi}\dot{\psi})+ \cos\theta (\psi' \chi' + \dot{\psi}\dot{\chi})=0=T_{\sigma \sigma}
\label{e17}
\end{eqnarray}
and,
\begin{eqnarray}
T_{\tau \sigma}=T_{\sigma \tau}=-r \dot{t}t' +\frac{\dot{r}r'}{4 r^2} +r \dot{\varphi}\varphi' +\frac{1}{4}(\dot{\chi}\chi' +\dot{\theta}\theta' +\dot{\psi}\psi')+ \lambda r( \dot{t}\chi' +\dot{\chi}t')\nonumber\\
+\lambda r \cos\theta (\dot{t}\psi' +\dot{\psi}t') + \lambda r (\dot{\varphi}\chi' + \dot{\chi}\varphi')+\lambda r \cos\theta (\dot{\varphi}\psi' +\dot{\psi}\varphi')+\frac{\cos\theta}{2}(\dot{\chi}\psi' + \dot{\psi}\chi')=0.
\label{e18}
\end{eqnarray}
\subsubsection{1D reduction: $ R \times S^3 $}
Our next step would be to \emph{consistently} reduce the parent 2D sigma model (\ref{e12}) into one dimension and thereby study the resulting phase space dynamics in the (semi)classical limit. In order to do so, we choose to work with the following string embedding,
\begin{eqnarray}
t =\kappa \tau ~;~ r=r_0 = const.~;~ \varphi =const.~;~ \psi = \psi(\tau) ~;~ \theta = \theta (\tau)~;~\chi = \chi (\tau)
\label{e19}
\end{eqnarray}
which, without loss of any generality, assumes the stringy dynamics over warped $ R \times S^3 $ subspace of the full target space geometry.

The Lagrangian density of the truncated theory could be formally expressed as,
\begin{eqnarray}
\mathfrak{L}_{1D}=\kappa^2 r_0 -\frac{1}{4}(\dot{\theta}^2 + \dot{\chi}^2 +\dot{\psi}^2)
- 2 \kappa \lambda  r_0 \dot{\chi} 
-2\kappa \lambda r_0 \cos\theta \dot{\psi}-\cos\theta \dot{\psi}\dot{\chi}.
\label{e20}
\end{eqnarray} 

The resulting equations of motion could be readily obtained as,
\begin{eqnarray}
\label{e21}
\ddot{\theta}+4 \kappa \lambda r_0 \sin\theta \dot{\psi}+2\sin\theta \dot{\psi}\dot{\chi} &=&0\\
\label{e22}
\ddot{\chi}+2\cos\theta \ddot{\psi}-2\sin\theta \dot{\theta}\dot{\psi}&=&0\\
\ddot{\psi}+2 \cos\theta \ddot{\chi}-4 \kappa \lambda r_0 \sin\theta \dot{\theta}-2\sin\theta \dot{\theta}\dot{\chi}&=&0.
\label{e23}
\end{eqnarray}

Our next task would be to check the Virasoro constraints (\ref{e17}) and (\ref{e18}) such that the time evolutions for them vanish when (\ref{e21})-(\ref{e23}) are implemented. This would eventually prove the consistency of our stringy ansatz (\ref{e19}). 

A straightforward computation reveals,
\begin{eqnarray}
T_{\tau \tau}&=& T_{\sigma \sigma}=- r_0 \kappa^2 +\frac{1}{4}(\dot{\chi}^2 + \dot{\theta}^2+\dot{\psi}^2)+2\kappa \lambda r_0 \dot{\chi}+2 \kappa \lambda  r_0 \cos\theta \dot{\psi}+\cos\theta \dot{\psi}\dot{\chi} \\
T_{\tau \sigma}&=& T_{\sigma \tau}=0.
\end{eqnarray}

Using (\ref{e21})-(\ref{e23}), and after a tedious but straightforward computation one finds,
\begin{eqnarray}
\partial_{\tau}T_{\tau \tau}=\kappa \lambda r_0 \ddot{\chi}.
\label{e26}
\end{eqnarray}\\
$ \bullet $ \textbf{Case I:} In order for the R.H.S. of (\ref{e26}) to be equal to zero, we must therefore set either $ r_0=0 $ or $ \ddot{\chi}=0 $. However, the first choice has to be ruled out as it does not see the effects of deformations on the boundary CFT$ _2 $ and the theory becomes a trivial sigma model on $ S^3 $ in the presence on NS-NS fluxes. Hence the consistency requirement of the Virasoro constraints sets, $ \dot{\chi}=-\lambda \kappa =$ constant. Also we set, $ r_0 =\frac{1}{2}$ without any loss of generality.

On the other hand, the conserved charges ($ \mathcal{Q}_i~;~i=1 ,2, 3 $) associated with the reduced sigma model (\ref{e20}) turns out to be,
\begin{eqnarray}
E&=&\frac{1}{2\pi \alpha'}\int_{0}^{2\pi}d\sigma \mathcal{E}=\frac{1}{\alpha'}(\kappa + \lambda^2 \kappa -\lambda \cos\theta \dot{\psi})\\
J_{\chi}&=&\frac{1}{2\pi \alpha'}\int_{0}^{2\pi}d\sigma \mathcal{J}_{\chi}=-\frac{1}{\alpha'}( \frac{\lambda \kappa}{2}  + \cos\theta \dot{\psi})\\
J_{\psi}&=&\frac{1}{2\pi \alpha'}\int_{0}^{2\pi}d\sigma \mathcal{J}_{\psi}=-\frac{\dot{\psi}}{2\alpha'}.
\end{eqnarray}

The consistency requirements, $ \partial_{\tau}\mathcal{Q}_i =0 $ further amounts of setting,
\begin{eqnarray}
\label{e30}
\cos\theta \ddot{\psi}-\sin\theta \dot{\theta}\dot{\psi}&=&0\\
\ddot{\psi}& =&0.
\label{e31}
\end{eqnarray}

It is trivial to see that the L.H.S. of (\ref{e22})-(\ref{e23}) vanish identically once (\ref{e30}) and (\ref{e31}) are implemented. Taking into account all the above mentioned facts, the reduced phase space dynamics boils down to the following single equation,
\begin{eqnarray}
\ddot{\theta} =0
\label{e32}
\end{eqnarray}
where, we set $ \dot{\psi}=\pm \lambda \kappa $ without loss of any generality. 

The above equation (\ref{e32}) has to be understood as the reduced phase space equation subjected to the constraints, $ J_{\psi}= $constant and $ J_{\chi}= $ constant where we focus on the dynamics over $ \lbrace \theta , J_{\theta}\sim \dot{\theta}\rbrace $ plane. Here, $ J_{\theta} $ is the momentum conjugate to $ \theta $.  The invariant plane \cite{Stepanchuk:2012xi} in this reduced phase space may be obtained by setting, $ \theta = \dot{\theta}=0 $ which trivially solves (\ref{e32}) and therefore is an allowed space of solution. 

In order to obtain the so called normal variational equation (NVE) one therefore needs to consider fluctuations over this invariant $\lbrace \theta=0 $, $ J_{\theta}=0\rbrace $ plane of solutions namely, $ \delta \theta \sim \eta (\tau) $ and retain upto leading order in the fluctuations which thereby yields,
\begin{eqnarray}
\ddot{\eta} \approx 0.
\label{e33}
\end{eqnarray}

Clearly, the NVE (\ref{e33}) allows Liouvillian solution of the form,
\begin{eqnarray}
\eta (\tau)\sim \mathcal{D} \tau + \tilde{\mathcal{D}}
\end{eqnarray}
where, $ \mathcal{D} $ and $ \tilde{\mathcal{D}} $ are arbitrary constants.

Looking back at (\ref{e10}) one finds $ V(\tau)=0 $ and therefore,
\begin{eqnarray}
w(\tau)\sim \frac{1}{\tau}
\end{eqnarray}
as a rational polynomial of degree 1 which clearly passes the Kovacic's test \cite{kovacic1}-\cite{kovacic2} and ensures the analytic integrability of the associated 1D sigma model.\\\\
$ \bullet $ \textbf{Case II:} The second possibility that we wish to explore is to set, $ \dot{\chi}=\lambda \kappa $. This results in the conserved charges of the following form,
\begin{eqnarray}
\label{E36}
E&=&\frac{1}{2\pi \alpha'}\int_{0}^{2\pi}d\sigma \mathcal{E}=\frac{1}{\alpha'}(\kappa - \lambda^2 \kappa -\lambda \cos\theta \dot{\psi})\\
J_{\chi}&=&\frac{1}{2\pi \alpha'}\int_{0}^{2\pi}d\sigma \mathcal{J}_{\chi}=-\frac{1}{\alpha'}( \frac{3\lambda \kappa}{2}  + \cos\theta \dot{\psi})\\
J_{\psi}&=&\frac{1}{2\pi \alpha'}\int_{0}^{2\pi}d\sigma \mathcal{J}_{\psi}=-\frac{1}{2\alpha'}(\dot{\psi}+4 \kappa \lambda \cos\theta).
\label{E38}
\end{eqnarray}

The conservation of (\ref{E36})-(\ref{E38}) yields the following set of constraints,
\begin{eqnarray}
\label{E39}
\cos\theta \ddot{\psi}-\sin\theta \dot{\theta}\dot{\psi}&=&0\\
\ddot{\psi}-4 \kappa \lambda \sin\theta \dot{\theta}& =&0.
\label{E40}
\end{eqnarray}

Notice that, the L.H.S. of (\ref{e22}) and (\ref{e23}) identically vanish once (\ref{E39}) and (\ref{E40}) are implemented. Effectively, the resulting dynamics therefore boils down to the following set of equations,
\begin{eqnarray}
\label{E41}
\ddot{\theta}+4 \kappa \lambda \sin\theta \dot{\psi}&=&0\\
\ddot{\psi}-4 \kappa \lambda \sin\theta \dot{\theta}& =&0.
\label{E42}
\end{eqnarray}

Clearly, $ \ddot{\theta}=\dot{\theta}=\theta=0 $ is a solution to (\ref{E41}) and may be used to set as an invariant plane \cite{Stepanchuk:2012xi} for the phase space with, $ \lbrace \theta=0, J_{\theta}\sim \dot{\theta}=0\rbrace $. Substituting this condition into (\ref{E42}) we set,
\begin{eqnarray}
\dot{\psi}=\lambda \kappa =const.
\label{E43}
\end{eqnarray}
without any loss of generality.

Using (\ref{E43}) and considering fluctuations, $ \delta \theta \sim \eta (\tau) $ normal to the invariant plane considered above, we arrive at the following NVE
\begin{eqnarray}
\ddot{\eta}+4 \kappa^2 \lambda^2 \eta \approx 0
\end{eqnarray}
which allows a Liouvillian form of solution,
\begin{eqnarray}
\eta (\tau)\sim \mathfrak{a}\cos n\tau + \mathfrak{b}\sin n\tau ~;~ n=2 \kappa \lambda
\end{eqnarray}
where $ \mathfrak{a} $ and $ \mathfrak{b} $ are two arbitrary constants.

Looking back at (\ref{e10}), we notice that the corresponding value for $ V(\tau)=-4\kappa^2 \lambda^2 $ which thereby yields,
\begin{eqnarray}
w_{(\pm)}\sim \frac{1}{4} \left(-\tau ^2 \pm 2 c_1 \tau -c_1^2+4 \kappa ^2 \lambda ^2\right)
\end{eqnarray}
which is a polynomial of degree 2 and therefore ensures that the associated phase space and hence the parent sigma model is classically integrable.
\subsubsection{1D reduction: $ BTZ \times S^1 $}
We now choose to work with the reduced sigma model by setting the ansatz,
\begin{eqnarray}
t =\kappa \tau ~;~ r=r(\tau)~;~ \varphi =\varphi (\tau)~;~ \psi = const. ~;~ \theta = const.~;~\chi = \chi (\tau)
\label{e36}
\end{eqnarray}
which considers string motion over warped $ BTZ \times S^1 $ subspace of the full 6D geometry.

The resulting Lagrangian density turns out to be,
\begin{eqnarray}
\mathfrak{L}_{1D}=\kappa^2 r - \frac{\dot{r}^2}{4r^2}-r \dot{\varphi}^2 - \frac{\dot{\chi}^2}{4} - 2 \kappa \lambda r \dot{\chi}-2\lambda r \dot{\varphi}\dot{\chi}.
\label{e37}
\end{eqnarray}

The equations of motion that readily follow from (\ref{e37}) could be formally expressed as,
\begin{eqnarray}
\label{e38}
r \ddot{r} - \dot{r}^2 -2r^3 (\dot{\varphi}^2 + 2\kappa \lambda \dot{\chi}+2\lambda \dot{\varphi}\dot{\chi}-\kappa^2)&=&0\\
\label{e39}
r \ddot{\varphi}+\dot{r}\dot{\varphi}+\lambda (\dot{r}\dot{\chi}+r \ddot{\chi})&=&0\\
\ddot{\chi}+4\kappa \lambda \dot{r}+4\lambda (\dot{r}\dot{\varphi}+r \ddot{\varphi})&=&0.
\label{e40}
\end{eqnarray}

The Virasoro constraints (\ref{e17})-(\ref{e18}) for the reduced sigma model turns out to be,
\begin{eqnarray}
T_{\tau \tau}&=& T_{\sigma \sigma}=-\kappa^2 r +\frac{\dot{r}^2}{4r^2}+r \dot{\varphi}^2 +\frac{\dot{\chi}^2}{4}+2 \kappa \lambda r \dot{\chi}+2\lambda r \dot{\varphi}\dot{\chi} \\
T_{\tau \sigma}&=& T_{\sigma \tau}=0.
\end{eqnarray}

Like before, we perform the consistency check of the stringy reduction by computing,
\begin{eqnarray}
\partial_{\tau}T_{\tau \tau}=-2 \kappa^2 \dot{r}+ 2\kappa \lambda \dot{r}\dot{\chi}+2 \kappa \lambda r \ddot{\chi}=0
\label{e43}
\end{eqnarray}
which we treat as an additional dynamical constraint of the system.

On the other hand, the conserved charges are estimated to be,
\begin{eqnarray}
E &=&\frac{1}{\alpha'}( 2\kappa r -2 \lambda r \dot{\chi})\\
P_{\varphi}&=&-\frac{1}{\alpha'}(2 r\dot{\varphi}+2 \lambda r \dot{\chi})\\
J_{\chi}&=&-\frac{1}{2\alpha'}(\dot{\chi}+4 \kappa \lambda r + 4 \lambda r \dot{\varphi}).
\end{eqnarray}

Notice that, $ \partial_{\tau}E=-\kappa^{-1}\partial_{\tau}T_{\tau \tau} =0 $. On the other hand, $  \partial_{\tau} P_{\varphi}=0$ by virtue of (\ref{e39}) where as the conservation of $ J_{\chi} $ follows directly from (\ref{e40}).

Combining (\ref{e38})-(\ref{e40}) and (\ref{e43}) we are finally left with the following set of dynamical equations,
\begin{eqnarray}
\label{e47}
r \ddot{r} - \dot{r}^2 -2r^3 (\dot{\varphi}^2 + 2\kappa \lambda \dot{\chi}+2\lambda \dot{\varphi}\dot{\chi}-\kappa^2)&=&0\\
\label{e48}
r \ddot{\varphi}+\dot{r}\dot{\varphi}+\kappa \dot{r}&=&0\\
\ddot{\chi}+4\kappa \lambda \dot{r}+4\lambda (\dot{r}\dot{\varphi}+r \ddot{\varphi})&=&0.
\label{e49}
\end{eqnarray}

In order to obtain NVE in a consistent way, we set $ r=r_0= $ constant which chooses the invariant plane \cite{Stepanchuk:2012xi} in the phase space as, $\lbrace r=r_0 , \mathcal{J}_r \sim \dot{r}=0 \rbrace$. Substituting this into (\ref{e48}) and (\ref{e49}) we find,
\begin{eqnarray}
\ddot{\varphi}=\ddot{\chi}=0
\end{eqnarray}
which thereby amounts of setting, 
\begin{eqnarray}
\dot{\varphi}=\dot{\chi}=\lambda \kappa = const.
\label{e51}
\end{eqnarray}
without any loss of generality. 

Substituting (\ref{e51}) into (\ref{e47}) we find the following constraint,
\begin{eqnarray}
2 \lambda^3 + 3 \lambda^2 -1=0
\end{eqnarray}
on the deformation parameter ($ \lambda $) which is trivially solved for, $ \lambda = \frac{1}{2} $.

Taking into account all the above inputs, we finally consider fluctuations,
\begin{eqnarray}
r \sim r_0 + \eta (\tau)
\end{eqnarray}
and arrive at the following normal variational equation, 
\begin{eqnarray}
\ddot{\eta}\approx 0
\end{eqnarray}
which yields Liouvillian solution of the form,
\begin{eqnarray}
\eta (\tau)\sim \mathfrak{a} \tau +\mathfrak{b}
\end{eqnarray}
where $ \mathfrak{a} $ and $ \mathfrak{b} $ are two arbitrary constants.

Comparing with (\ref{e8}) we further notice that $ \mathcal{A}=1 $ together with $ \tilde{\mathcal{B}}=\tilde{\mathcal{C}}=0 $ which thereby sets,
\begin{eqnarray}
w'(\tau)+w^2(\tau)=V(\tau)=0.
\label{e56}
\end{eqnarray}
The solution to the above equation (\ref{e56}) is a rational polynomial $ w(\tau)\sim \frac{1}{\tau} $ of degree 1 thereby clearly fulfilling the Kovacic's criteria. In summary, the above analysis ensures the classical integrability of strings over warped $ BTZ \times S^1 $.
\subsection{Sigma models on warped BTZ black string }
We now start considering sigma models over warped BTZ black string background \cite{Apolo:2019yfj},
\begin{eqnarray}
ds^2 =\frac{f(r)}{4}dr^2 +r (d \varphi^2 -dt^2)+\ell (\lambda)r (d\varphi + dt)(d\chi +\cos\theta d\psi)~~~~~~~~~~~~~~\nonumber\\
+2T^2_v \ell (\lambda)(d\varphi - dt)(d\chi +\cos\theta d\psi)+T^2_u (d\varphi + dt)^2+T^2_v (d\varphi - dt)^2+d\Omega^{2}_3
\end{eqnarray}
where we identify each of the individual entities as,
\begin{eqnarray}
f(r)=\frac{1}{r^2 - r_t^2}~;~r^2_t = 4T^2_u T^2_v~;~\ell (\lambda)=\frac{\lambda T_v}{1+ \lambda^2 T^2_v}
\end{eqnarray}
with $ T_{u,v} $ as some dimensionless temperatures related to the left and right moving energies of the BTZ black string \cite{Azeyanagi:2012zd}.

The NS-NS sector following the TsT reads as \cite{Apolo:2019yfj},
\begin{eqnarray}
\mathcal{B}_2 =\frac{\cos\theta}{4} d\psi \wedge d \chi +r d \varphi \wedge dt -\frac{r}{2}\ell (\lambda)(d \varphi +dt)\wedge(d\chi + \cos\theta d\psi) \nonumber\\
- T^2_v \ell (\lambda)(d\varphi - dt)\wedge (d\chi + \cos\theta d\psi)+\frac{\lambda}{2}T^2_v (d\varphi - dt)\wedge d\chi.
\end{eqnarray}
\subsubsection{1D reduction: $ R \times S^3 $}
Like before, our next step would be to consistently reduce the parent 2D sigma model over warped $ R \times S^3 $ and study the resulting phase space dynamics in the (semi)classical limit. In order to do so, we choose to work with the following ansatz,
\begin{eqnarray}
t =\kappa \tau ~;~ r=r_0 = const.~;~ \varphi =const.~;~ \psi = \psi(\tau) ~;~ \theta = \theta (\tau)~;~\chi = \chi (\tau).
\label{e60}
\end{eqnarray}

The resulting Lagrangian density turns out to be,
\begin{eqnarray}
\mathfrak{L}_{1D}=\beta \kappa^2 -\frac{1}{4}(\dot{\theta}^2 +\dot{\chi}^2 +\dot{\psi}^2)-2 \kappa \tilde{\ell}\dot{\chi}-2\kappa \tilde{\ell}\cos\theta \dot{\psi}-\cos\theta \dot{\chi}\dot{\psi}
\end{eqnarray}
where we define,
\begin{eqnarray}
\beta = r^2_0 -T_u^2-T_v^2~;~\tilde{\ell}(\lambda)=(r_0 - 2 T^2_v)\ell (\lambda).
\end{eqnarray}

The corresponding equations of motion could be formally expressed as,
\begin{eqnarray}
\label{e63}
\ddot{\theta}+4 \kappa \tilde{\ell} \sin\theta \dot{\psi}+2\sin\theta \dot{\psi}\dot{\chi} &=&0\\
\label{e64}
\ddot{\chi}+2\cos\theta \ddot{\psi}-2\sin\theta \dot{\theta}\dot{\psi}&=&0\\
\ddot{\psi}+2 \cos\theta \ddot{\chi}-4 \kappa\tilde{\ell} \sin\theta \dot{\theta}-2\sin\theta \dot{\theta}\dot{\chi}&=&0.
\label{e65}
\end{eqnarray}

The Virasoro constraints on the other hand turn out to be,
\begin{eqnarray}
\label{e66}
T_{\tau \tau}&=& T_{\sigma \sigma}=- \beta \kappa^2 +\frac{1}{4}(\dot{\chi}^2 + \dot{\theta}^2+\dot{\psi}^2)+2\kappa \tilde{\ell} \dot{\chi}+2 \kappa \tilde{\ell} \cos\theta \dot{\psi}+\cos\theta \dot{\psi}\dot{\chi} \\
T_{\tau \sigma}&=& T_{\sigma \tau}=0.
\end{eqnarray}

A straightforward computation reveals,
\begin{eqnarray}
\partial_{\tau}T_{\tau \tau}=\kappa \tilde{\ell}\ddot{\chi}
\end{eqnarray}
which amounts of setting, $ \ddot{\chi}=0 $ in order for (\ref{e66}) to be a valid constraint condition.\\\\
$ \bullet $ \textbf{Case I:} We note down conserved charges associated with the reduced sigma model,
\begin{eqnarray}
\label{e69}
E &=&\frac{2}{\alpha'}(\kappa \beta + \kappa \tilde{\ell}^2 -\tilde{\ell}\cos\theta \dot{\psi} )\\
J_{\chi}&=&-\frac{1}{\alpha'}(\frac{3}{2}\kappa \tilde{\ell}+\cos\theta \dot{\psi})\\
J_{\psi}&=&-\frac{1}{2\alpha'}(\dot{\psi}+2 \kappa \tilde{\ell}\cos\theta)
\label{e71}
\end{eqnarray}
where, we set $ \dot{\chi}=-\kappa \tilde{\ell}= $ constant without any loss of generality.

The conservation of (\ref{e69})-(\ref{e71}) yields the following set of constraint equations,
\begin{eqnarray}
\label{e72}
\cos\theta \ddot{\psi}-\sin\theta \dot{\theta}\dot\psi &=&0\\
\ddot{\psi}-2\kappa\tilde{\ell}\sin\theta \dot{\theta}&=&0.
\label{e73}
\end{eqnarray}

It is in fact interesting to note that the L.H.S. of (\ref{e64}) and (\ref{e65}) identically vanish once (\ref{e72}) and (\ref{e73}) are implemented. Therefore the final set of equations boil down to,
\begin{eqnarray}
\label{e74}
\ddot{\theta}+2 \kappa \tilde{\ell} \sin\theta \dot{\psi} &=&0\\
\ddot{\psi}-2\kappa\tilde{\ell}\sin\theta \dot{\theta}&=&0.
\label{e75}
\end{eqnarray}

Like before, we set $ \ddot{\theta}=\dot{\theta}=\theta =0 $ which trivially solves (\ref{e74}) and sets the invariant plane in the phase space as, $ \lbrace \dot{\theta}=0, J_{\theta}\sim \dot{\theta}=0 \rbrace$. Substituting this into (\ref{e75}) we find,
\begin{eqnarray}
\dot{\psi}= -\kappa \tilde{\ell}=\dot{\chi}=const.
\label{e76}
\end{eqnarray}

Using (\ref{e76}) and considering fluctuations $ \delta \theta \sim \eta (\tau) $ we arrive at the NVE of the following form,
\begin{eqnarray}
\ddot{\eta}-2\kappa^2 \tilde{\ell}^2 \eta \approx 0
\end{eqnarray}
which admits Liouvillian solution of the form,
\begin{eqnarray}
\eta (\tau)\sim e^{-\sqrt{2}\kappa \tilde{\ell}\tau}.
\end{eqnarray}

Looking back at (\ref{e10}) we find, $ V(\tau)=-\tilde{\mathcal{C}}=2\kappa^2 \tilde{\ell}^2 $ which thereby yields,
\begin{eqnarray}
w_{(\pm)}(\tau)\sim \frac{1}{4} \left(-\tau ^2\pm 2 c_1 \tau -c_1^2+8 \kappa ^2 \tilde{\ell}^2\right)
\end{eqnarray}
both of which are rational polynomials of degree 2.

A second possibility is to set, $ \dot{\psi}=\kappa \tilde{\ell}=-\dot{\chi} $ which yields a different NVE
\begin{eqnarray}
\ddot{\eta}+2\kappa^2 \tilde{\ell}^2 \eta \approx 0
\end{eqnarray}
that results in a Liouvillian solution of the form,
\begin{eqnarray}
\eta (\tau)\sim \mathfrak{a} \cos m \tau + \mathfrak{b} \sin m\tau ~;~ m=\sqrt{2}\kappa \tilde{\ell}.
\end{eqnarray}

Looking back at (\ref{e10}), we further notice that,
\begin{eqnarray}
w_{(\pm)}(\tau)\sim \frac{1}{4} \left(-\tau ^2\pm 2 c_1 \tau -c_1^2-8 \kappa ^2 \tilde{\ell}^2\right)
\end{eqnarray}
both of which are again polynomials of degree 2. All these results point towards the underlying integrable structure of the parent sigma model over warped $ R \times S^3 $ at finite temperature.\\\\
$ \bullet $ \textbf{Case II:} The second possibility that one might wish to explore is to set, $ \dot{\chi}=\kappa \tilde{\ell} $ which yields the following set of conserved charges,
\begin{eqnarray}
\label{e83}
E &=&\frac{2}{\alpha'}(\kappa \beta - \kappa \tilde{\ell}^2 -\tilde{\ell}\cos\theta \dot{\psi} )\\
J_{\chi}&=&-\frac{1}{\alpha'}(\frac{5}{2}\kappa \tilde{\ell}+\cos\theta \dot{\psi})\\
J_{\psi}&=&-\frac{1}{2\alpha'}(\dot{\psi}+6 \kappa \tilde{\ell}\cos\theta).
\label{e85}
\end{eqnarray}

The conservation of (\ref{e83})-(\ref{e85}) yields,
\begin{eqnarray}
\label{e86}
\cos\theta \ddot{\psi}-\sin\theta \dot{\theta}\dot\psi &=&0\\
\ddot{\psi}-6\kappa\tilde{\ell}\sin\theta \dot{\theta}&=&0
\label{e87}
\end{eqnarray}

Like before, the L.H.S of (\ref{e64}) and (\ref{e65}) identically vanish once (\ref{e86}) and (\ref{e87}) are implemented. Therefore, the dynamics finally boils down to the following set of equations,
\begin{eqnarray}
\label{e88}
\ddot{\theta}+6 \kappa \tilde{\ell} \sin\theta \dot{\psi} &=&0\\
\ddot{\psi}-6\kappa\tilde{\ell}\sin\theta \dot{\theta}&=&0
\label{e89}
\end{eqnarray}
which except for some numerical factors, look exactly identical to those obtained in the earlier example in (\ref{e74})-(\ref{e75}). Therefore without getting into further details, one can conclude that the parent string theory is classically integrable.
\subsubsection{1D reduction: $ BTZ \times S^1 $}
Finally, we choose to work with the string embedding,
\begin{eqnarray}
t =\kappa \tau ~;~ r=r(\tau)~;~ \varphi =\varphi (\tau)~;~ \psi = const. ~;~ \theta = const.~;~\chi = \chi (\tau)
\label{e101}
\end{eqnarray}
corresponding to warped $ BTZ \times S^1 $.

The corresponding Lagrangian density turns out to be,
\begin{eqnarray}
\mathfrak{L}_{1D}=\zeta (r)\kappa^2 -\frac{f(r)}{4}\dot{r}^2 - \frac{\dot{\chi}^2}{4}-\upsilon (r)\dot{\varphi}^2 -\tilde{\mu} \kappa \dot{\varphi}-2 \ell \varrho_{+}(r)\dot{\varphi}\dot{\chi}-2\kappa \ell \varrho_{-}(r)\dot{\chi}
\label{e102}
\end{eqnarray}
where we identify each of the above entities as,
\begin{eqnarray}
\tilde{\mu }&=& 4(T^2_u -T^2_v)~;~\varrho_{\pm}(r)=r \pm 2T^2_v\\
\zeta (r)&=&r-T^2_u - T^2_v ~;~\upsilon(r)=r+T^2_u +T^2_v.
\end{eqnarray}

Below we note down equations of motion that readily follows from (\ref{e102})
\begin{eqnarray}
\label{e105}
2 f(r)\ddot{r}+f'(r)\dot{r}^2- 4 (\dot{\varphi}^2 +2\ell \dot{\varphi}\dot{\chi}+2 \kappa \ell \dot{\chi}-\kappa^2)&=&0\\
\upsilon (r)\ddot{\varphi}+\dot{r}\dot{\varphi}+\ell (\dot{r}\dot{\chi}+\varrho_{+}(r)\ddot{\chi})&=&0\\
\ddot{\chi}+4\kappa \ell \dot{r}+4 \ell (\dot{r}\dot{\varphi}+\varrho_{+}(r)\ddot{\varphi})&=&0.
\label{e107}
\end{eqnarray}

The Virasoro constraints, on the other hand, are given by
\begin{eqnarray}
T_{\tau \tau}=T_{\sigma \sigma}=-\zeta (r)\kappa^2 +\frac{f(r)}{4}\dot{r}^2 + \frac{\dot{\chi}^2}{4}+\upsilon (r)\dot{\varphi}^2 \nonumber\\
+\tilde{\mu} \kappa \dot{\varphi}+2 \ell \varrho_{+}(r)\dot{\varphi}\dot{\chi}+2\kappa \ell \varrho_{-}(r)\dot{\chi}
\end{eqnarray}
and,
\begin{eqnarray}
T_{\tau \sigma}=T_{\sigma \tau}=0.
\end{eqnarray}

A straightforward computation reveals,
\begin{eqnarray}
\partial_{\tau}T_{\tau \tau}=-2 \kappa^2 \dot{r}+2\kappa \ell \dot{r}\dot{\chi}+2\kappa \ell \varrho_{-}(r)\ddot{\chi}+\tilde{\mu}\kappa \ddot{\varphi}=0
\label{e110}
\end{eqnarray}
which we take as an additional dynamical constraint of our system.

The conserved charges for the system on the other hand turns out to be,
\begin{eqnarray}
\label{e111}
E &=&\frac{1}{\alpha'}( 2\kappa \zeta (r)-\tilde{\mu}\dot{\varphi} -2 \ell \varrho_{-}(r) \dot{\chi})\\
P_{\varphi}&=&-\frac{1}{\alpha'}(2 \upsilon (r)\dot{\varphi}+\tilde{\mu}\kappa +2 \ell \varrho_{+}(r) \dot{\chi})\\
J_{\chi}&=&-\frac{1}{2\alpha'}(\dot{\chi}+4 \kappa \ell \varrho_{-}(r) + 4 \ell \varrho_{+}(r) \dot{\varphi}).
\label{e113}
\end{eqnarray}
Like before, the conservation of (\ref{e111})-(\ref{e113}) is ensured once (\ref{e105})-(\ref{e107}) together with (\ref{e110}) are implemented.

Therefore, finally we are left with the following set of equations,
\begin{eqnarray}
\label{e114}
2 f(r)\ddot{r}+f'(r)\dot{r}^2- 4 (\dot{\varphi}^2 +2\ell \dot{\varphi}\dot{\chi}+2 \kappa \ell \dot{\chi}-\kappa^2)&=&0\\
\label{e115}
(\upsilon (r)-\frac{\tilde{\mu}}{2})\ddot{\varphi}+\dot{r}\dot{\varphi}+4 T^2_v\ell\ddot{\chi}+\kappa \dot{r} &=&0\\
\ddot{\chi}+4\kappa \ell \dot{r}+4 \ell (\dot{r}\dot{\varphi}+\varrho_{+}(r)\ddot{\varphi})&=&0.
\label{e116}
\end{eqnarray}

In order to define the invariant plane in the phase space, we first project ourselves to the subspace of the full dynamical phase space where both $ P_{\varphi}=J_{\chi}= $ constant which amounts of setting, $ \dot{\chi}=\dot{\varphi}=\kappa \ell $ together with, $ r=r_0 $ without any loss of generality. The above choice clearly solves (\ref{e115}) and (\ref{e116}) and enforces us to set, $ \ell = \frac{1}{2} $ in order to satisfy (\ref{e114}). This makes the above choice as the allowed space of solution for the dynamical phase space under consideration. Finally, considering fluctuations, $ r \sim r_0 + \eta (\tau) $ and retaining upto linear order, we arrive at the NVE 
\begin{eqnarray}
\ddot{\eta}\approx 0
\end{eqnarray}
which allows Liouvillian form of solution as found earlier. This also ensures the corresponding polynomial function, $ w(\tau)\sim \frac{1}{\tau} $ to be of degree 1 which is again consistent with Kovacic's criteria and therefore ensures the integrability of the underlying phase space.
\section{Summary and final remarks}
We show analytic integrability for classical strings probing warped $ BTZ\times S^3 $ geometries both for the zero as well as finite temperature examples. Our approach is based on Kovacic's algorithm that essentially allows one to check whether a particular phase space configuration is integrable or not. We reduce the parent sigma model over different subspaces of the full target space geometry and obtain the corresponding dynamical equations those are consistent with various other physical constraints of the sigma model. We then consider fluctuations over some invariant plane in the phase space which results in the normal variational equation (NVE). Furthermore, we show that the solution to the NVE is consistent with rules set by the algorithm.  This ensures the analytic integrability of classical strings that are dual to single trace operators in $ J\bar{T} $ deformed CFT$ _2 $.

The results of this paper could have a similar interpretation in terms of integrability preservation as in the case of dual CFT$ _2 $ in the presence of irrelevant deformations \cite{Anous:2019osb}-\cite{Conti:2019dxg}. The dual CFT$ _2 $ is solvable in the sense that the spectrum of the deformed theory could be expressed as an \emph{exact} function of the deformation parameter, the scale associated with the problem and the original (undeformed) CFT data \cite{Guica:2019vnb}-\cite{Guica:2017lia}. On the other hand, the Kovacic's algorithm essentially is a prescription of showing classical integrability of the dynamical phase space under consideration. Moreover, a state of \emph{fixed} energy ($ \Delta $) in the dual CFT$ _2 $ corresponds to stringy excitation ($ E_s $) in terms of sigma model degrees of freedom and therefore can be mapped to a fixed energy orbit in the dynamical phase space. Putting all these pieces together, a natural interpretation would therefore be that in the presence of marginal deformations, the fixed energy orbits in the phase space could be obtained precisely by knowing the original phase space configuration of the sigma (WZW) model in the undeformed theory. In other words, the integrable trajectories (at fixed energy) of the deformed theory could be generated in terms of the original (undeformed) phase space data as well as by knowing the value of the deformation parameter.\\\\
{\bf {Acknowledgements :}}
 The author is indebted to the authorities of IIT Roorkee for their unconditional support towards researches in basic sciences. 

\end{document}